\documentclass[twocolumn,aps,floatfix,showpacs,showkeys]{revtex4}
\usepackage{color,graphicx}% Include figure files
\usepackage{dcolumn}% Align table columns on decimal point
\usepackage{bm}% bold math
\usepackage{epsfig}
\usepackage{supertabular}
\usepackage[bookmarksopen]{hyperref}
\def\be {\begin{equation}}
\def\ee {\end{equation}}
\def\nn {\nonumber}
\def\bea {\begin{eqnarray}}
\def\eea {\end{eqnarray}}

\begin{document}
\title{Dijet induced collective modes in an anisotropic quark-gluon plasma
}
\bigskip
\bigskip
\author{ Mahatsab Mandal}
\email{mahatsab.mandal@saha.ac.in}
\author{ Pradip Roy}
\email{pradipk.roy@saha.ac.in}
\affiliation{Saha Institute of Nuclear Physics, 1/AF Bidhannagar
Kolkata - 700064, India}

\begin{abstract}
We discuss the collective modes due to the propagation of two oppositely
moving relativistic 
jets (dijet) in an anisotropic quark-gluon plasma(AQGP) and compare the results
with the case of single jet propagation.
%The plasma and jets of particles can be described by Vlasov-type transport equations.
%To simplify our analysis we assume that 
For the sake of simplicity, assuming 
a tsunami-like initial jet distribution, 
we observe that the dispersion relations for both the stable and 
unstable modes are altered significantly 
due to the passage of dijet in comparison with the case of single 
jet propagation. 
It has been further demonstrated that
the growth rate of instability, due to
introduction of dijet in the system, increases compared 
to the case of single jet case. As in the case of single jet 
propagation, the instability always grows when the jet velocity is 
perpendicular to the wave vector.  
We, thus,  argue that the introduction
of dijet in the AQGP, in general, leads to faster isotropization than
single jet propagation.
 
%for the special case when the wave vector is parallel to the anisotropy axis.
%We calculate the stable and unstable modes for the (...........) and two stream plasma. We find that ......  
\end{abstract}
\keywords{collective mode, anisotropy, relativistic jets}
\pacs{12.38.Mh, 05.20.Dd}

\maketitle

%% main text
\section{Introduction}
One of the goals of the ultra-relativistic heavy-ion collision 
experiments at BNL RHIC and at CERN LHC is 
to produce a deconfined state of QCD matter commonly
known as quark gluon plasma (QGP).
The properties of such a system, if formed, can
be studied through various probes such as electromagnetic probe,
jet quenching etc..
According to QCD thermodynamics studied on lattice, this
 novel state of matter is 
expected to be formed at temperatures of the order of $T\sim170 - 200$MeV. 
The jet of particles produced initially due to hard collision, when passes
through such a system loses energy and results in decrease of
high energy hadrons produced due to the jet fragmentation.
This phenomenon is commonly known as jet quenching.
Moreover, the passage of jets influences
the collective modes of the system~\cite{prd74,prd76,prd77,maha1,prd81}.  

%One of the aspects of the ongoing ultra-relativistic heavy-ion collision experiments at BNL RHIC and the 
%experimental at CERN LHC are to study the behavior of the possible plasma phase of QCD, the so-called quark-gluon plasma (QGP).
%High energy  partons behaves as hard probes which are produced in the early stage of collision due to hard 
%scattering. When the jet of partons passes through the hot and dense medium it losses  energy, mainly by 
%radiative processes. This phenomena is commonly known as jet quenching, because in the direction of propagation 
%of the jet one observed a decrease of high energy hadrons and increase in the number of soft hadrons. 

%Various models have been proposed to describe the jet quenching mechanism~\cite{jetquen}. 
The effect of single jet propagation on the plasma collective modes
has been studied in Refs~\cite{prd74,prd76,prd77} where it has been assumed
that the plasma is isotropic in momentum space and the jet is
described by a tsunami-like jet distribution. 
In a very recent work the present authors have relaxed the
condition of isotropic QGP and studied the characteristics
of both stable and unstable collective modes of AQGP due to the passage of
relativistic jet using transport theory~\cite{maha1}. 
It is shown~\cite{maha1} that the dispersion
relations are modified quite substantially in comparison to the no
jet case. 
On the other hand, the authors of Ref~\cite{prd81}, have
investigated the dijet induced properties of the
unstable modes in isotropic QGP.
% due to the passage of
%dijet. 
Also the energy loss due to stream 
instabilities induced by dijet has been discussed using both tsunami-like
and Gaussian distributions for the relativistic jets within the
same approach as in Refs.~\cite{prd74,prd76,prd77,maha1}. 
We extend the work of Ref.~\cite{prd81,maha1} to AQGP and study the
characteristics of both stable and unstable modes due to the
passage of two jets moving in opposite direction.
It should be noted that non-equilibrium jet of particles while travelling 
through QGP disturbs the plasma exciting chromomagnetic
and chromoelectric modes of which some of modes might even be unstable. 
The most important among those are the modes which grow exponentially in time. 
Plasma instabilities due to jet propagation
in AQGP have been the subject of extensive studies as
when the instability occurs, the kinetic energy of the  particles is 
converted to the field energy  leading to faster 
isotropization~\cite{prc49,prl94,prl94A,appb} 
of the QGP. This type of phenomenon occurs if it is assumed that the 
the hadronization time is greater than the
time to generate the growth of the gauge fields.
For the complete development of plasma instabilities, 
the time scale should be of the order of 
$t\sim(6.7-12.5)/\omega_t$, where 
$\omega_t$ is the total angular frequency of the whole 
system~\cite{prd76,prd77}. 

We use the same simplifying assumptions as in Ref.~\cite{prd76,prd77,prd81,maha1}
to study the stable and unstable modes of AQGP induced by two 
counter propagating 
jets and compare it with that obtained in case of single jet~\cite{maha1}
and no jet case~\cite{prd68, prd70}.
AQGP is realized in the very early stages of heavy-ion collision due to rapid 
longitudinal expansion. As a consequence, the cooling is faster
in the longitudinal direction which results in 
$\langle p_L^2\rangle<<\langle p_T^2\rangle$. Such momentum-space 
anisotropy modifies the  collective modes which have different
behavior in comparison to that in isotropic QGP 
(see Refs.~~\cite{prd68, prd70} for details).  
Due to the passage of relativistic jets in a non-equilibrium  
plasma, with an anisotropic distribution in momentum space, the behavior of 
collective modes  change as will be demonstrated in the following. 

The essential ingredient to study the collective modes in the composite system
(AQGP+jets) is the transport equation which reduces to Vlasov 
equations for the time scale shorter than the mean free path 
time~\cite{prc49,pr183}.
In our calculation we assume (i) weak coupling regime i.e. $g << 1$ 
and (ii) the interactions between jet and 
plasma is only mediated by mean gauge fields. 
The interactions between the jet particles and the plasma particles
leads kinetic instability initiated either by charge or current 
fluctuations which lead to electric and magnetic instabilities.
The latter is  also known as Weibel instability~\cite{prl2}. 
%In  the first case, the electric field is longitudinal 
%i.e. the field is parallel to the wave vector ${\bf k}$ (${\bf E}||{\bf k}$), 
%while in the second case the field is perpendicular to ${\bf k}$ (${\bf E} \perp {\bf k}$). 
%For this reason the corresponding instabilities are called longitudinal and 
%transverse instabilities respectively. 
%Since the electric field plays a crucial role in the generation of longitudinal 
%modes, they are also called electric modes, while the transverse modes are 
%called magnetic modes. The magnetic mode known as filamentation- or 
%Weibel-instability~\cite{prl2} appears to be relevant for the QGP~\cite{prc49,appb}. 
It has been demonstrated in Ref.~\cite{prd68,prd70} that 
the growth rate of magnetic 
instability is maximum in the direction of anisotropy  
in momentum-space anisotropic plasma. 
So while studying the characteristics of the unstable modes
we assume that the momentum of the 
collective mode is in the direction of the anisotropy. 

The organization of the paper is as follows. In section 2 we  
mention the required formalism very briefly as detail has already been
presented in \cite{prd76,prd77,prd81,maha1}.
In section 3 results for both stable and unstable modes
will be presented followed by summary in section 4.

%The plan of the paper is as follows. In section 2 we briefly describe the Vlasov-type transport equation and 

\section{Formalism}
To study the jet induced collective modes in AQGP we dwell on the
transport theory described in detail in Ref.~\cite{prd76,prd77,prd81,maha1}.
For the sake of simplicity, we assume two jets propagating in opposite
direction described by the following tsunami-like distribution~\cite{prd77}
\be
f_{jet}({\bf p})={\bar n}{\bar u}^0\delta^{(3)}({\bf p}-\Lambda {\bar {\bf u}}).\label{td}
\ee
Here ${\bar n}$ is related to the density of the plasma 
and ${\bar u}^{\mu}$ 
is the four-velocity. The parameter $\Lambda$ fixes the scale of 
energy of the particles. More realistic distribution, such as, a Gaussian 
distribution can be used to simulate the same effect~\cite{prd81}.

Using Vlasov approximation one can write down the polarization tensor 
for particles species $\alpha$ as~\cite{prd76,prd77,prd68,prd73}:
\be
\Pi^{\mu\nu}_{\alpha}(k)=g^2\int_p p^{\mu}\frac{\partial f_{\alpha}(\bf p)}{\partial p_{\beta}}
\Big(g^{\beta\nu}-\frac{p^{\nu}k^{\beta}}{p.k+i\epsilon}\Big)\label{dt}
\ee
where $\alpha$ specify the quarks, antiquarks, gluons or partials of jet. 
The polarization tensor is symmetric, i. e. $\Pi^{\mu\nu}(k)=\Pi^{\nu\mu}(k)$,
and transverse, $k^{\mu}\Pi^{\mu\nu}=0$.

We assume that the ansatz for the anisotropic momentum distribution is
given by~\cite{prd68,prd70}:
\be
f({\bf p})=f_{\xi}({\bf p})=N(\xi)f_{iso}(\sqrt{{\bf p}^2+\xi({\bf p.\hat{n}})^2}).
\ee
Here $f_{iso}$ is the  distribution function in isotropic case. 
$N(\xi)$ is the normalization constant which is equal to $\sqrt{1+\xi}$,
${\bf \hat{n}}$ is the unit vector along the direction of anisotropy. 
The anisotropy parameter $\xi$ lies in the range $-1<\xi<\infty$.
Using the above ansatz the spacelike component of the self-energy tensor can be written as (from Eq.(\ref{dt})): 
\be
\Pi^{ij}_p(k)=m_D^2\sqrt{1+\xi}\int \frac{d\Omega}{(4\pi)}\frac{v^l+\xi({\bf v.\hat{n}})n^l}{(1+\xi({\bf v.\hat{n}})^2)^2}
\Big(\delta^{jl}+\frac{v^jk^l}{(k.v+i\epsilon}\Big)
\ee
where
\be
m_D^2=-\frac{g^2}{2\pi^2}\int^{\infty}_0dpp^2\frac{df_{iso}(p^2)}{dp}
\ee
is the isotropic Debye mass.
Since the self energy depends on the four-momentum($k^{\mu}$) and the 
the anisotropic vector $(n^{\mu}=(1,{\bf n}))$,
it can be decomposed in a suitable tensorial form in a covariant 
gauge~\cite{prd68,plb662}:
\be
\Pi^{ij}_p(k)=\alpha A^{ij}+\beta B^{ij}+\gamma C^{ij}+\delta D^{ij}
\ee
where 
\bea A^{ij}&=&\delta^{ij}-k^ik^j/{\bf k^2},\label{s1}\\
B^{ij}&=&k^ik^j/{\bf k^2},\label{s2}\\
C^{ij}&=&\tilde{n}^i\tilde{n}^j/\tilde{n}^2,\label{s3}\\
D^{ij}&=&k^i\tilde{n}^j+k^j\tilde{n}^i,\label{s4}
\eea
where $\tilde{n}^i=A^{ij}n^j$ which obeys ${\bf\tilde{n}.k}=0$
and the structure functions $\alpha, \beta, \gamma$ and $\delta$ can
be obtained by appropriate 
contractions (see \cite{prd68,plb662} for details).
%\bea
%k^i\Pi^{ij}k^j&=&{\bf k}^2\beta,\nonumber\\
%\tilde{n}^i\Pi^{ij}k^j&=&\tilde{n}^2{\bf k}^2\delta,\nonumber\\
%\tilde{n}^i\Pi^{ij}\tilde{n}^j&=&\tilde{n}^2(\alpha+\gamma),\nonumber\\
%{\rm Tr}\Pi^{ij}&=&2\alpha+\beta+\gamma
%\eea
In the limit of $\xi\rightarrow0,$ (isotropic case)
$\gamma$ and $\delta$ vanish whereas, $\alpha$ and $\beta$ are 
directly related to the transverse and longitudinal components of 
the polarization tensor of the plasma respectively.

In a similar way, 
the polarization tensors induced 
by the jets (moving in opposite direction)
with the tsunami-like momentum distribution 
can be calculated to obtain the following expressions~\cite{prd81}:
\bea
\Pi^{ij}_{jet1}(k)=-\omega^2_{jet}\Big(\delta^{ij}
+\frac{k^iv^j_{jet}+k^jv^i_{jet}}{\omega-{\bf k.v}_{jet}}-
\frac{(\omega^2-{\bf k}^2)v_{jet}^iv_{jet}^j}
{(\omega-{\bf k.v}_{jet})^2}\Big),\nn\\
\eea
\bea
\Pi^{ij}_{jet2}(k)=-\omega^2_{jet}\Big(\delta^{ij}-
\frac{k^iv^j_{jet}+k^jv^i_{jet}}{\omega+{\bf k.v}_{jet}}-
\frac{(\omega^2-{\bf k}^2)v_{jet}^iv_{jet}^j}
{(\omega+{\bf k.v}_{jet})^2}\Big),\nn\\
\eea
where $v_{jet}$ is the velocity of jet and 
$\omega^2_{jet}=\frac{g^2\bar{n}}{2\Lambda}$ is the plasma frequency of the jet. 
Moreover, the polarization tensors  due to the jets can be 
decomposed in the following way:
\be
\Pi^{ij}_{jet~1} = \alpha^{\prime}A^{ij} + \beta^{\prime}B^{ij} 
\ee
\be
\Pi^{ij}_{jet~2} = \alpha^{\prime\prime}A^{ij} + \beta^{\prime\prime}B^{ij} 
\ee
where
\bea
\alpha^{\prime}&=&-\frac{\omega_{jet}}{2}[2-\frac{\omega^2-k^2}
{(\omega-kv_{jet}\cos\theta_{jet})^2}v^2\sin^2\theta_{jet}]\nn\\
\beta^{\prime}&=&\omega^2_{jet}\omega^2\frac{v^2\cos^2\theta_{jet}-1}
{(\omega-kv_{jet}\cos\theta_{jet})^2}
\eea
and
\bea
\alpha^{\prime\prime}&=&-\frac{\omega_{jet}}{2}[2-\frac{\omega^2-k^2}
{(\omega+kv_{jet}\cos\theta_{jet})^2}v^2\sin^2\theta_{jet}]\nn\\
\beta^{\prime\prime}&=&\omega^2_{jet}\omega^2\frac{v^2\cos^2\theta_{jet}-1}
{(\omega+kv_{jet}\cos\theta_{jet})^2}.
\eea
For numerical convenience, we introduce another parameter $\eta$ 
defined by $\eta=\omega_{\rm jet}^2/\omega_t^2$ where $\omega_t^2=
\omega_{\rm jet}^2+m_D^2/3$.

%The dispersion laws of the collective modes of the 
%system due to dijet are  determined by searching the poles of 
%the propagator of Eq.(\ref{pole}) by replacing $\Pi^{ij}_p$ with $\Pi^{ij}_{jet1}$ 
%and $\Pi^{ij}_{jet2}$ i.e. by finding the solution $\omega(k)$. 

In the analysis of the collective modes of the composite system
we are interested in very short time scales where the collisionless
approximation is justified. The effect of jet of particle is to induce
the color fluctuations, which provide a contribution to the polarization
tensor of the system. According to the linear response theory, 
the total polarization tensor of the whole system is given by the sum 
of polarization tensors of the plasma and the two jets.
Thus we have,
\be
\Pi^{\mu\nu}_t(k)=\Pi^{\mu\nu}_{p}(k)+\Pi^{\mu\nu}_{jet1}(k)+\Pi^{\mu\nu}_{jet2}(k)\label{pit}
\ee  
The dispersion relation of the collective modes of  the total system can be 
determined by solving the equation
\be
{\rm det}[(k^2-\omega^2)\delta^{ij}-k^ik^j+\Pi^{ij}_{t}(k)]={\rm det}[\Delta^{-1}(k)]^{ij}=0.\label{dr}
\ee
where, in the temporal axial gauge,  the effective propagator $\Delta^{ij}$ is
given by,
\be
\Delta^{ij}(k)=\frac{1}{[({\bf k}^2-\omega^2)\delta^{ij}-k^ik^j+\Pi_t^{ij}(k)]}.\label{pole}
\ee
Therefore, using Eqs.(6), (11), (12), and (17) the effective propagator 
of the composite system can be written as 
\bea
\Delta(k)&=&\Delta_A{\bf A}+({\bf k}^2-\omega^2+\alpha+
\alpha^{\prime}+\alpha^{\prime\prime}+\gamma)\Delta_G{\bf B}\nn\\
&+&[(\beta+\beta^{\prime}+\beta^{\prime\prime}-
\omega^2)\Delta_G-\Delta_A]{\bf C}-\delta\Delta_G{\bf D}
\eea
with
\bea
\Delta^{-1}_A &=& {\bf k}^2-\omega^2+\alpha+
\alpha^{\prime}+\alpha^{\prime\prime},\label{c1}\\
\Delta^{-1}_G&=&({\bf k}^2-\omega^2+\alpha+\alpha^{\prime}+
\alpha^{\prime\prime}+\gamma)(\beta+\beta^{\prime}+
\beta^{\prime\prime}-\omega^2)\nonumber\\
&-&{\bf k}^2\tilde{n^2}\delta^2.\label{c2}
\eea

\begin{figure}[t]
\begin{center}
\epsfig{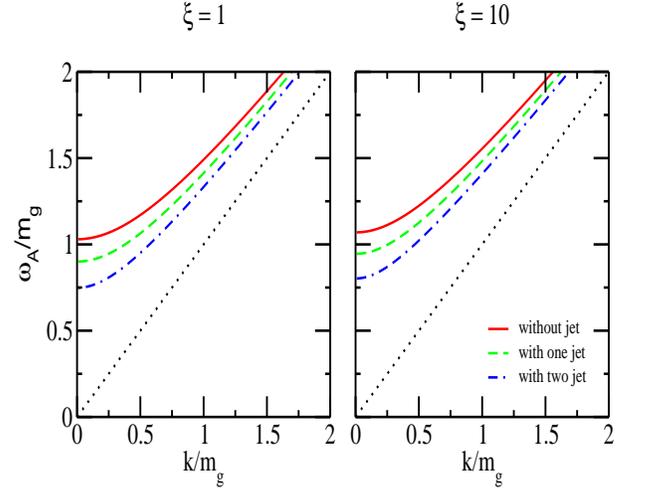}
\end{center}
\caption{(Color online) The dispersion relation for the stable A-mode for an anisotropic plasma 
with jet for different values of the anisotropy parameter $\xi=\{1, 10\}$, $\eta=0.2$, $\theta_{jet}=0$,
$v_{jet}=0.7$ and $\theta_n=0$}  
\label{fig1}
\end{figure}

\begin{figure}[t]
\begin{center}
\epsfig{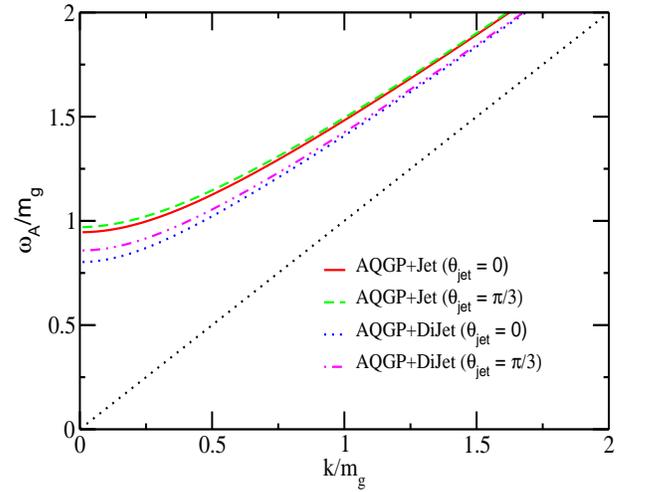}
\end{center}
%\caption{(Color online) Same as Fig.\protect\ref{fig1}
\caption{(Color online) The dispersion relation for the stable A-mode for the system 
composed of anisotropic plasma with single jet and dijet 
for  $\xi=10$, $v_{jet}=0.7$, $\theta_{n}=0$, $\eta=0.2$ and $\theta_{jet}=\{0,~\pi /3\}$.}  
\label{fig2}
\end{figure}

\begin{figure}[t]
\begin{center}
\epsfig{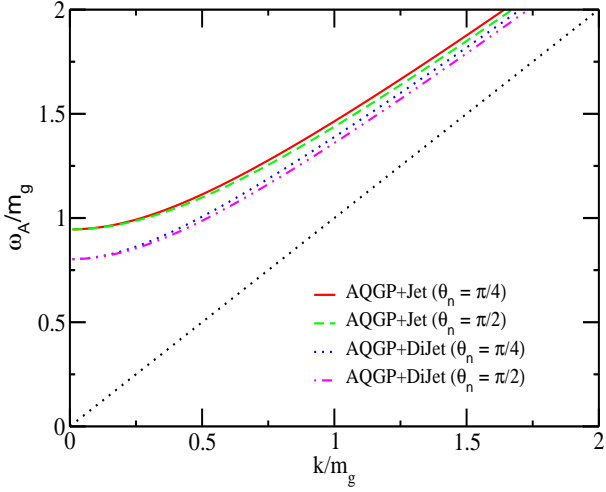}
\end{center}
\caption{(Color online) Same as Fig.\protect\ref{fig2} for
different values of $\theta_n=\{\pi/4,~\pi/2\}$ and $\xi=10$, $\theta_{jet}=0$, $\eta=0.2$ and $v_{jet}=0.7$.} 
\label{fig3}
\end{figure}
\begin{figure}[t]
\begin{center}
\epsfig{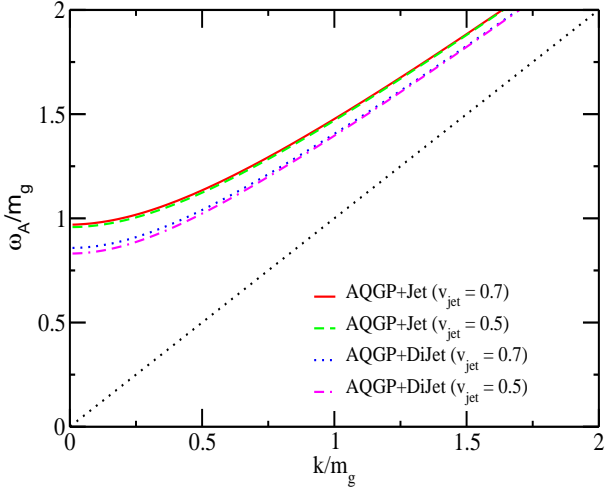}
\end{center}
\caption{(Color online) Same as Fig.\protect\ref{fig2} for 
$\xi=10$, $\theta_n=\pi/4$ $\theta_{\rm jet}=\pi/3$, $\eta=0.2$ and  $v_{\rm jet}=\{0.5,~0.7\}$} 
\label{fig4}
\end{figure}

\section{Results}
\subsection{Stable modes}
The collective modes of the system comprising of the plasma and the
jets are obtained by finding the zeros of Eqs. 
(\ref{c1}) and (\ref{c2}).
For real values of $\omega>|{\bf k}|$, the dispersion relation for $A$-modes can be determined by 
finding the solutions of the equation:
\bea
\omega^2_A = {\bf k}^2+\alpha(\omega_A)+\alpha^{\prime}(\omega_A)+\alpha^{\prime\prime}(\omega_A)~\label{as}
\eea
In case of $G$-modes we factorize $\Delta_G^{-1}$ as
\bea
\Delta_G^{-1}=(\omega^2-\omega^2_{G+})(\omega^2-\omega^2_{G-})
\eea
where 
\bea
\omega^2_{G\pm} = \frac{1}{2}(\bar{\omega}^2\pm\sqrt{\Omega^2+
4{\bf k}^2\tilde{n^2}\delta^2}),~\label{cm2}
\eea
and 
\bea
\bar{\omega}^2 = \alpha+\alpha^{\prime}+\alpha^{\prime\prime}+\beta+\beta^{\prime}+\beta^{\prime\prime}+\gamma+{\bf k}^2,\nonumber\\
\Omega=\alpha+\alpha^{\prime}+\alpha^{\prime\prime}-\beta-\beta^{\prime}-\beta^{\prime\prime}+\gamma+{\bf k}^2.~\label{cm3}
\eea
For real $\omega>|{\bf k}|$, the square root of Eq. (\ref{cm2}) is always 
positive leading to two stable modes for $G$-modes.

\begin{figure}[t]
\begin{center}
\epsfig{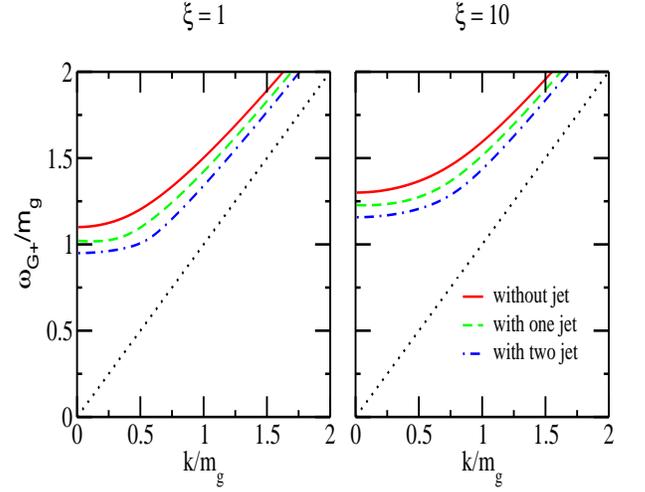}
\end{center}
\caption{(Color online) The dispersion relation for the stable $G_+$-mode for an anisotropic plasma 
with jet  for $\theta_n=\pi/4$, $\theta_{jet}=0$, $\eta=0.2$ and $v_{jet}=0.7$. 
The left(right) panel corresponds to anisotropy parameter $\xi=1(10)$.}  
\label{fig5}
\end{figure}
\begin{figure}[t]
\begin{center}
\epsfig{file=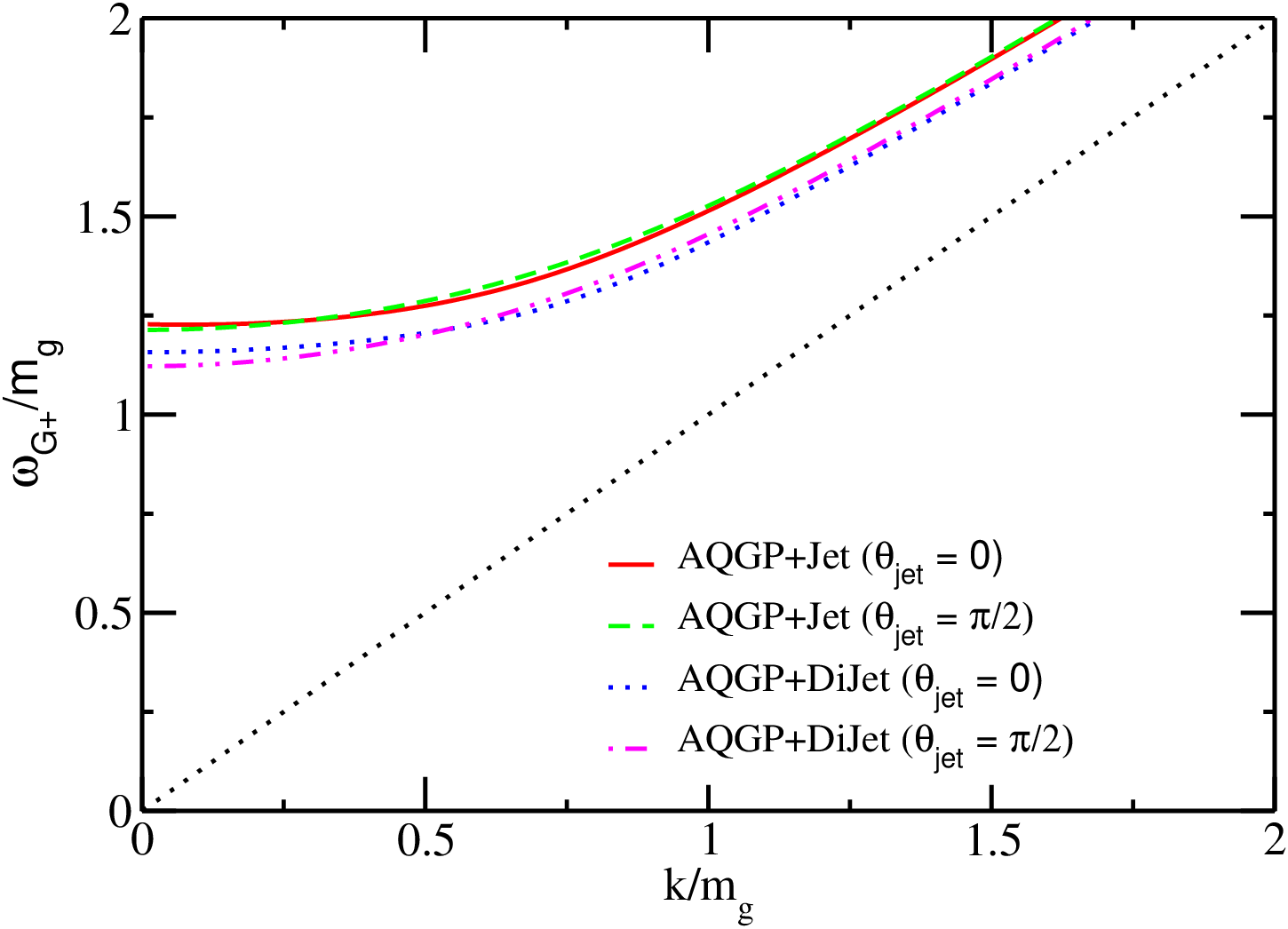,width=8cm,height=6.5cm,angle=0}
\end{center}
\caption{(Color online) The dispersion relation for the stable $G_+$-mode for the system composed of anisotropic plasma 
with single  jet and dijet for $\xi=10$, $v_{jet}=0.7$, $\theta_{n}=\pi/4$, $\eta=0.2$ and different $\theta_{jet}=\{0,~\pi /2\}$.}  
\label{fig6}
\end{figure}
\begin{figure}[t]
\begin{center}
\epsfig{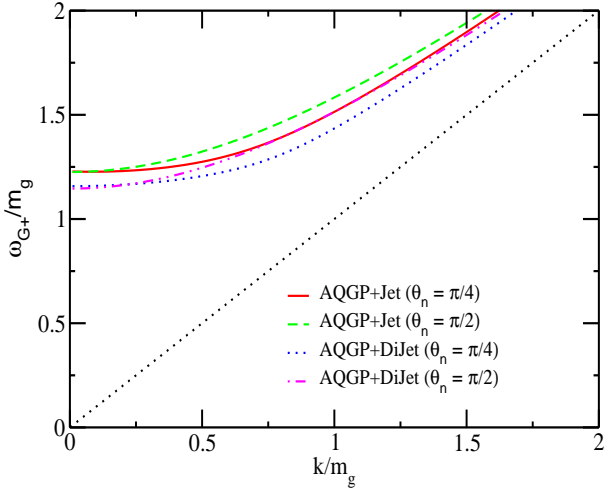}
\end{center}
\caption{(Color online) Same as Fig.\protect\ref{fig6}
%The dispersion relation for the stable $G_+$-mode for the system composed of anisotropic plasma and  jet 
for$\xi=10,$ $v_{jet}=0.7$, $\theta_{jet}=0$ and $\eta=0.2$ and $\theta_n=\{\pi/4,~\pi/2\}$.}  
\label{fig7}
\end{figure}
\begin{figure}[t]
\begin{center}
\epsfig{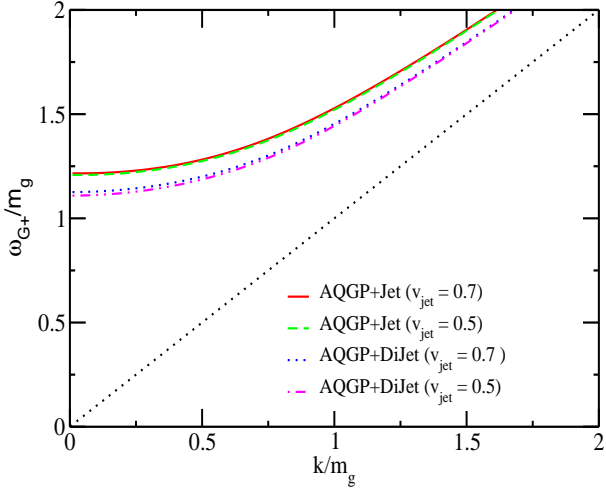}
\end{center}
\caption{(Color online)  Same as Fig.\protect\ref{fig6}
%The dispersion relation for the stable $G_+$-mode for the system composed by anisotropy plasma and a jet 
for $\theta_{n}=\pi/4$, $\xi=10,$ $\eta=0.2$, $\theta_{jet}=\pi/3$ and $v_{jet}=\{0.7,~0.5\}$.}  
\label{fig8}
\end{figure}

% figs. for \omega_G- modes
\begin{figure}[t]
\begin{center}
\epsfig{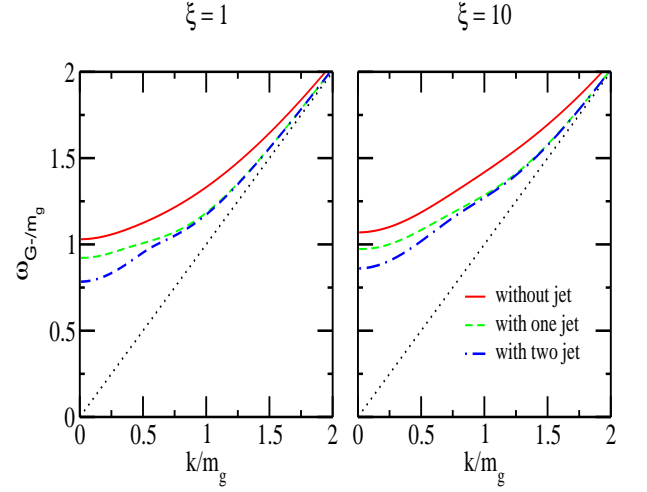}
\end{center}
\caption{(Color online) The dispersion relation for the stable $G_-$-mode for an anisotropy plasma and 
a jet for $\theta_n=\pi/4$, $\theta_{jet}=0$, $v_{jet}=0.7$ and $\eta=0.2$. 
The left(right) panel corresponds to anisotropy parameter $\xi=1(10)$.}  
\label{fig9}
\end{figure}

\begin{figure}[t]
\begin{center}
\epsfig{file=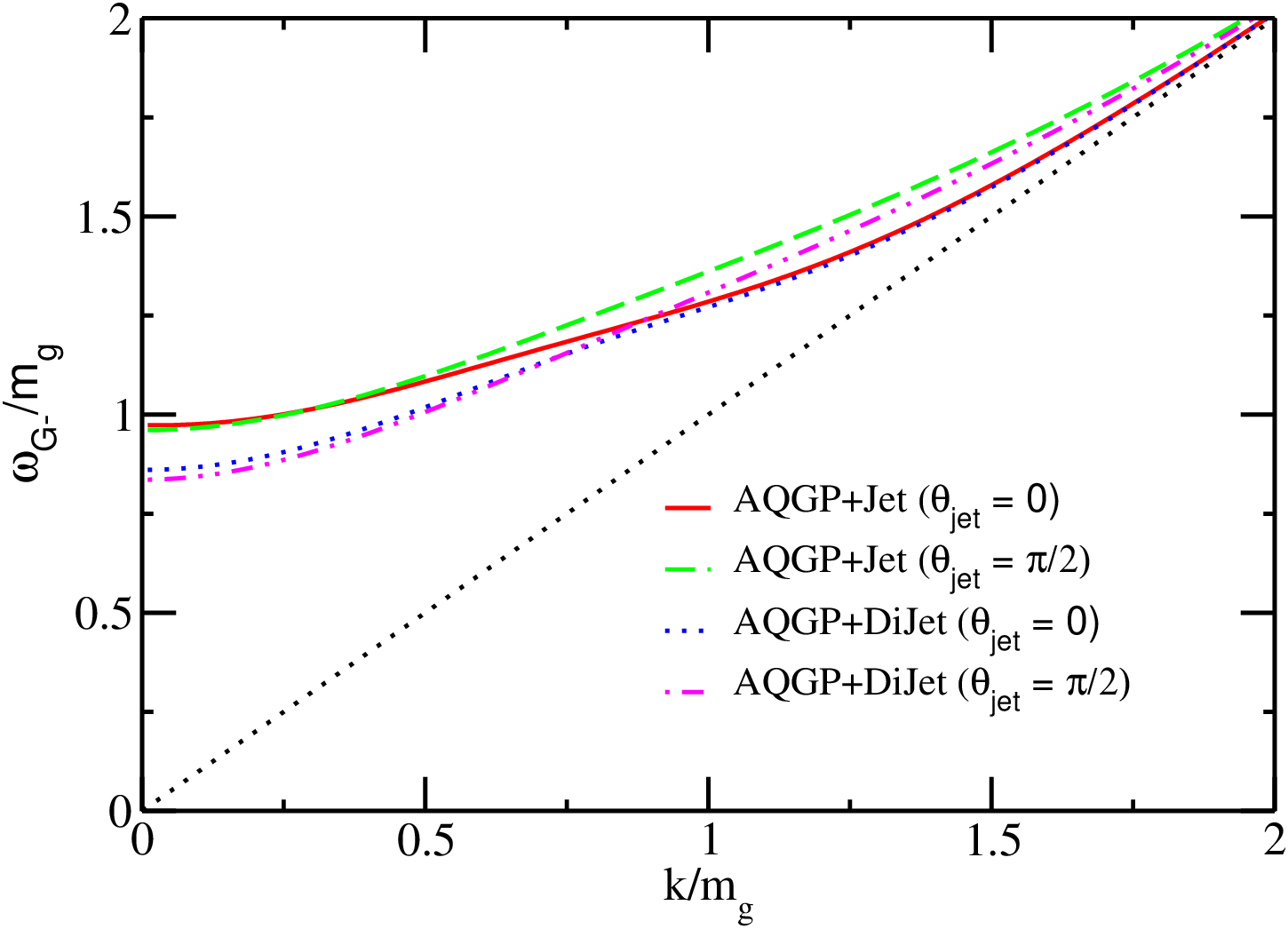,width=8cm,height=6.5cm,angle=0}
\end{center}
\caption{(Color online) The dispersion relation for the stable $G_-$(G)-mode 
for the composed of anisotropic plasma with single jet and dijet  
for  $\theta_{jet}=\{0,~\pi /2\}$, $\xi=10$, $v_{jet}=0.7$, $\theta_{n}=\pi/4$ and $\eta=0.2$.}  
\label{fig10}
\end{figure}
\begin{figure}[t]
\begin{center}
\epsfig{file=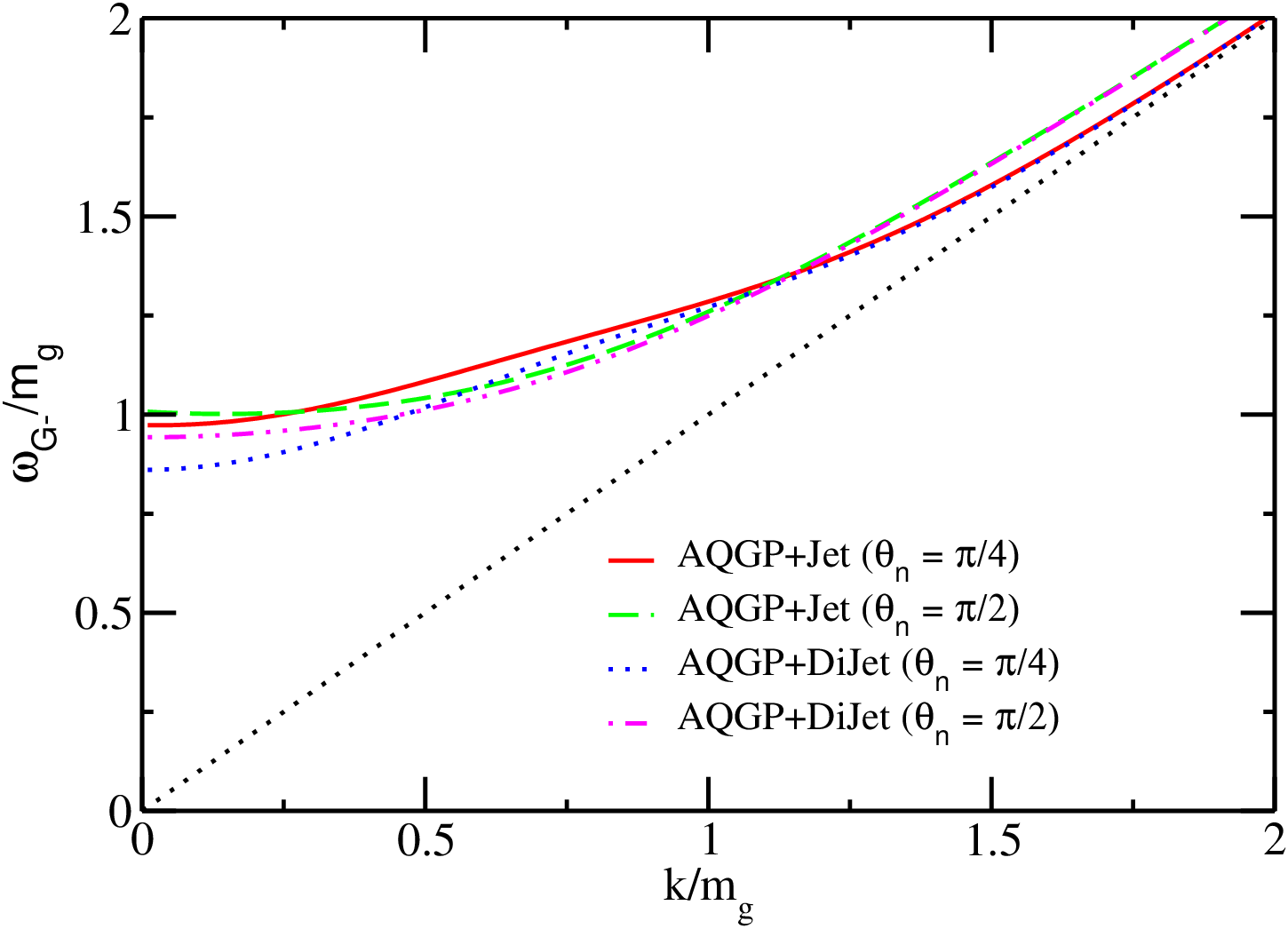,width=8cm,height=6.5cm,angle=0}
\end{center}
\caption{(Color online) Same as Fig.\protect\ref{fig10}
%The dispersion relation for the stable $G_-$-mode for the system composed by anisotropy plasma and a jet 
for $\xi=10,$ $v_{jet}=0.7$, $\theta_{jet}=0$, $\eta=0.2$ and different $\theta_n=\{\pi/4,~\pi/2\}$.}  
\label{fig11}
\end{figure}

\begin{figure}[t]
\begin{center}
\epsfig{file=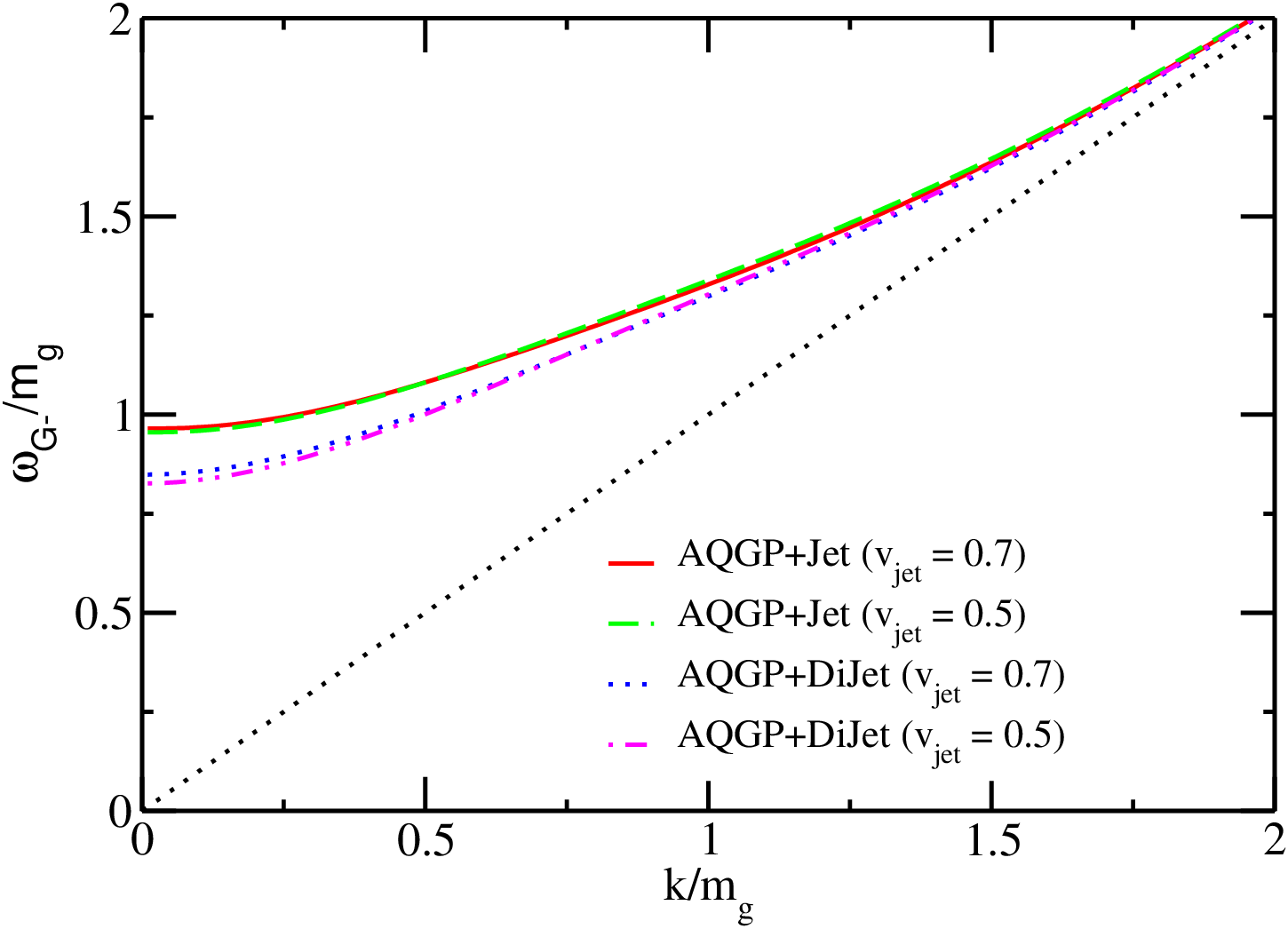,width=8cm,height=6.5cm,angle=0}
\end{center}
\caption{(Color online) Same as Fig.\protect\ref{fig10}
%The dispersion relation for the stable $G_-$-mode for the system composed by anisotropy plasma and a jet 
for $\xi=10$, $\theta_{n}=\pi/4$, $\theta_{jet}=\pi/3$, $\eta=0.2$ and different $v_{jet}=\{0.7,~0.5\}$.}  
\label{fig12}
\end{figure}

We vary various parameters introduced earlier to find the solutions 
numerically.
We, in  Fig.\ref{fig1}, plot the results for the stable $A$-mode with
single jet and dijet for various parameters explained in the figure. For
completeness we have also plotted the results when there is no jet.
%for two values of the anisotropic parameter $\xi=\{1,~10\}$, $\eta=0.2$, $\theta_{jet}=0$, 
%$v_{jet}=0.7$ and $\theta_{n}=0$ 
We see that the collective modes with the dijet differs reasonably
from that with single jet and also with no jet. 
In order to see the the dependence of the dispersion relation on 
the angle of propagation of the jet with the wave vector the results for
the $A$-modes have been 
displayed in Fig.\ref{fig2} for $\xi=10$, $v_{jet}=0.7$, $\theta_{n}=0$ 
and $\eta=0.2$ where one can clearly see that the dispersion
relation is sensitive to $\theta_{jet}$. Next we consider
the variation of the dispersion relations for $A$-modes
with the anisotropy direction
$\theta_n$. For fixed $\theta_{jet}$ the results are in shown
in Fig.\ref{fig3}.
The dispersion relation for dijet (jet) is not at all sensitive
at low momentum and is marginally sensitive at higher momentum. 
The important thing is to notice that the dijet results for
fixed $\theta_n$ substantially differ from that with the single jet.
The sensitivity of the dispersion relation for the stable A-modes 
on the jet velocity is displayed
in Fig.\ref{fig4} for $\theta_n=\pi/4$. 
We see that the results changes marginally by varying the
jet velocity both for single jet and dijet propagation.

%For fixed value of $\theta_n$ the modes have significant dependence on the value of $\eta$.
It can be checked that $\gamma$ and $\tilde{n}^2$ vanish identically when 
${\bf k}||{\bf \hat{n}}$. 
This leads to similar dispersion relations for $G_+$-modes for $\theta_n=0$.  
First of all, we display the results for $G_+$-modes 
in Fig.\ref{fig5} for two values of the anisotropic parameter $\xi$. 
The results are quite sensitive to $\xi$ for fixed $\theta_n$.
It is also seen that for fixed $\xi$ and $\theta_n$ the dijet results
differ substantially from the single jet case.
We also find marginal dependence of $G_+$-modes on $\theta_{jet}$ 
at very low momentum(see Fig.\ref{fig6}).
%The collective modes also depend on the jet 
%direction as depicted in Fig.\ref{fig6}. 
%However we do not find any significant dependence of the $G_+$-modes on the jet 
%velocity(see in Fig.\ref{fig6}).
In Fig.\ref{fig7} the sensitivity of the dispersion relation 
of the $G_+$-modes with the anisotropy direction ($\theta_n$) 
has been displayed. It 
is found that the stable $G_+$-modes are quite sensitive to
$\theta_n$. It is also observed that the dijet results are 
quite different from the single jet case which is an important 
finding in the present work. We have also found that as in the case 
of single jet, 
dijet results are not that sensitive to the jet velocity
(see Fig.\ref{fig8}).  
%We also find marginal dependence of G-modes on $\eta$ and $\theta_n$ (not shown in here).
Finally, for stable modes, we consider the $G_-$-modes which have
been shown in Fig.\ref{fig9} for  $\xi=\{1,~10\}$, $\eta=0.2$, 
$\theta_{jet}=0$, $v_{jet}=0.7$ and $\theta_{n}=\pi/4$ with and without jets. 
We find  significant difference in the 
collective modes between single jet and dijet. 
The $\theta_{\rm jet}$-dependence of the $G_-$-modes
for $\theta_{jet}=\{0,\pi/2\}$ is displayed in Fig.\ref{fig10}. 
It is seen that the dispersion relation for $G_-$-mode is more sensitive 
to $\theta_{jet}$ which is not the case for $A$ and $G_+$-modes, 
The dispersion relation for $G_-$-modes for various $\theta_n$ is plotted 
in Fig.\ref{fig11}. We find significant dependencies on $\theta_n$. 
The collective mode for $G_-$-mode is also 
marginally sensitive to the jet velocity(see Fig.\ref{fig12}). 
%In Figs.\ref{fig2} and \ref{fig4}, it is seen that very small dependence on the angle of propagation
%with respect to velocity of the particle of jet, $\theta_{jet}$ and velocity of particle of jet, $v_{jet}$ respectively. 
%From the Fig.\ref{fig3}, it is clearly  shown that the dispersion 
%relation depends on $\eta$ and the angle between  the propagation vector and the anisotropy vector, $\theta_{n}$.
%When the wave vector is parallel to the direction of the anisotropy, i.e., ${\bf k}||{\bf \hat{n}}$ then $\gamma$ and $\tilde{n}^2$
%vanish identically. Therefore  the dispersion relation for $G+$ modes are similar to $A$ modes.
%The dispersion relation for $G+$ modes are shown in Fig.\ref{fig5} for different $\xi=\{1,~10\}$, together with anisotropic 
%quark-gluon plasma and the anisotropy plasma induced by jet. The corresponding value of the $\theta_n=\pi/2$, 
%$\eta=0.2$, $\theta_{jet}=0$ and $v_{jet}=0.7$. In Figs.\ref{fig6},\ref{fig7} and \ref{fig8} are shown that dependence on 
%the angle $\theta_{jet}$, $v_{jet}$ and $\eta$ respectively. In  Fig.\ref{fig9} we plot the angular dependence of $\omega_{G+}$ 
%for $\xi=10$, $v_{jet}=0.7$, $\theta_{jet}=\pi/2$, $\eta=0.2$ and $\theta_{n}=\{0,~\pi/2,~\pi/4\}$.

\begin{figure}
\begin{center}
\epsfig{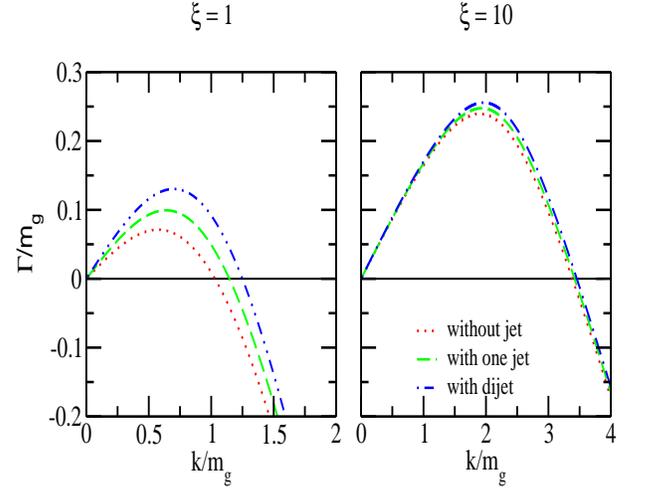}
\end{center}
\caption{(Color online) The growth rate $\Gamma$ of the unstable  mode of single jet and dijet for two different anisotropy parameter 
$\xi=\{1, 10\}$ and $\theta_{jet}=0$, $\eta=0.2$ and $v_{jet}=0.7$.}  
\label{figu1}
\end{figure}

%$\theta_{jet}=\{0, \pi/4, \pi/3, \pi/2\}$, $\xi=10$ and $\eta=0.2$.}
%\label{figu2}
%\end{figure}

\begin{figure}
\begin{center}
\epsfig{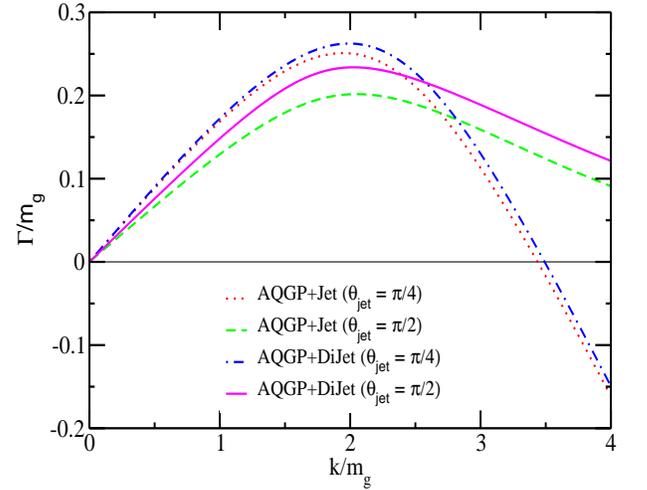}
\end{center}
\caption{(Color online) Same as Fig.\protect\ref{figu1} with 
$\xi=10$, $v_{jet}=0.7$, $\eta=0.2$ and $\theta_{jet}=\{\pi/4,~\pi/2\}$.}
\label{figu2}
\end{figure}

\subsection{Unstable modes}
In the static limit, for small $\theta_n$  
some of the scales appearing in the dispersion relations are negative implying 
that the  whole system is unstable with 
respect to magnetic  instability~\cite{bjp,jhep08}. 
It has been demonstrated that in case of 
${\bf k}\parallel{\bf \hat{n}}(\theta_n=0)$,  
the growth rate  of  the filamentation instabilities is  
the largest~\cite{prd68,prd70,prd73}.
In such case $\gamma$ and $\tilde{n}^2$ vanish identically and
the dispersion relations (henceforth called $\alpha$ and $\beta$ modes)
 for the unstable modes simplify to
\bea
\omega^2-{\bf k}^2-\alpha(\omega)-\alpha^{\prime}(\omega)-\alpha^{\prime\prime}(\omega)&=&0\\
\omega^2-\beta(\omega)-\beta^{\prime}(\omega)-\beta^{\prime\prime}(\omega)&=&0
\eea  
In the numerical solutions of the above equations
we have checked that unstable mode exits only 
for $\alpha$ mode both for single jet~\cite{maha1} and dijet. 
Let us now concentrate on the dispersion relations for the unstable $\alpha$ 
mode for which we shall, first assume that  $\theta_{jet}=0$.
It is numerically checked that $\omega$ is
purely imaginary, i.e. $\omega=i\Gamma$ with $\Gamma$ real valued. 
We first show our results for two values of the anisotropy parameter 
$\xi=\{1,~10\}$ and for fixed values of the other parameters in the
model. This is depicted in Fig.\ref{figu1}. 
The growth of the instability in case of dijet is more
compared to the single jet case for a fixed value of the anisotropy 
parameter. However, like the case of single jet results~\cite{maha1},
the instability first increases and then becomes damped. It is also seen
that the results are quite sensitive to $\xi$.
%To find the maximum value of the momentum $k_{max}$ at which the unstable mode spectrum terminates, we take the limit 
%$\Gamma\rightarrow0$, to obtain($\theta_{jet}=0$), 
%\bea
%k^2_{max} = \omega^2_{jet}+m_D^2\frac{\sqrt{\xi}+(\xi-1)\arctan(\sqrt{\xi})}{4\sqrt{\frac{\xi}{\xi+1}}}.
%\eea

For $\theta_{jet}\neq 0$ 
the roots of the dispersion relations are of the form
of $\omega=a+i\Gamma$, i. e. these are unstable propagating modes. 
This has been checked numerically. The results for the unstable
modes with non-zero $\theta_{\rm jet}$ is delineated in Fig.\ref{figu2}.
both for single jet and dijet. 
For $\theta_{\rm jet}=\pi/4$ the results are similar to the case of
$\theta_{\rm jet}=0$.
However, for $\theta_{\rm jet}=\pi/2$, the instability always grows 
for dijet case as 
well as in the case of single jet which has already been reported~\cite{maha1}.
In this case the modes  never become damped.  
It is also worthwhile to note that the dijet induced instability for
$\theta_n=\pi/2$ is more compared to the single jet case.

\section{Summary}
The dispersion relations of both stable and unstable collective modes in
AQGP due to the dijet propagation have been obtained and solved numerically.
We then compare it with that obtained in case of single jet and no jet cases. 
Depending upon the values of different parameters 
and for the dijet case considered here, the stable modes differ significantly
from that obtained with single jet. We have shown the dependencies 
of the dispersion relations (both for single and dijet) on various 
parameters explained in the text. 
It has been observed that the dispersion relation for stable modes are very
sensitive to the anisotropy parameter, the anisotropy direction and the jet 
direction with respect to the wave vector. For the unstable modes it is 
found that the growth rate with dijet is more compared to the single jet case. 
In particular, for $\theta_{jet}=\pi/2$, the growth rate for the dijet 
case is substantially larger and it always grows like the single 
jet case~\cite{maha1}. Thus the introduction of dijet might lead to 
faster isotropization as compared to the single jet case. For the special 
case considered here (${\bf{\hat k}}||{\bf\hat n}$) no unstable modes exits 
for $\beta$-mode even with the dijet. However, more general case might lead to
$\beta$-unstable modes which is worth investigating.

%We have also shown that the growth rate $\Gamma$ 
%increases with the passage of jet and this growth rate strongly depends on the anisotropy parameter $\xi$. This means that 
%the introduction of jet in an AQGP leads to faster isotropization for the special case $(\hat{k}||\hat{n})$ 
%considered here. It is also interesting to note 
%that the growth rate for $\theta_{jet}=\pi/2$ never becomes damped.
%We also find no unstable for $\beta-\beta^{\prime}$-mode  in the special case.
%However, it remains to be seen how the unstable modes behave when
%a more general case(arbitrary $\theta_n$) is considered. 

\end{document}